\documentclass[fleqn,usenatbib]{mnras}

\usepackage{newtxtext,newtxmath}

\usepackage[T1]{fontenc}

\DeclareRobustCommand{\VAN}[3]{#2}
\let\VANthebibliography\thebibliography
\def\thebibliography{\DeclareRobustCommand{\VAN}[3]{##3}\VANthebibliography}

\usepackage{graphicx}	
\usepackage{subcaption}
\usepackage{amsmath}	
\usepackage{hyperref}

\title[Orbital Retention of Habitable-Zone Planets]{Predicting the Orbital Retention of Habitable-Zone Planets Following Stellar Flybys Using Machine Learning}

\author[F. Rahimi and A. Farhani Asl]{
Fatemeh Rahimi$^{1}$\thanks{E-mail: f.rahimii2021@gmail.com},
Ahmad Farhani Asl$^{1}$
\\
$^{1}$Department of Physics, Institute for Advanced Studies in Basic Sciences (IASBS), Zanjan, Iran\\
}

\date{Accepted XXX. Received YYY; in original form ZZZ}

\pubyear{\the\year{}}

\hypersetup
{colorlinks=true, citecolor=blue, linkcolor=blue, urlcolor=blue}

\begin{document}
\markboth{F. Rahimi}{Shortened Paper Title}
\label{firstpage}
\pagerange{\pageref{firstpage}--\pageref{lastpage}}
\maketitle

\begin{abstract}
Stellar flybys can significantly alter the orbital configurations of planetary systems, potentially displacing planets initially located within the habitable zone (HZ) of their host stars. In this study, we investigate the dynamical effects of stellar encounters on HZ planets using a large suite of three-body N-body simulations. We examine how the properties of the host star, passing star, planetary orbit, and encounter geometry influence the final orbital configuration. A planet is considered to retain an HZ orbit only if it remains bound to its original host and its entire final orbital range, from periastron to apastron, lies within the HZ boundaries. Based on the simulation results, we develop an XGBoost classifier to predict the final HZ status of planets following stellar encounters. The model achieves an accuracy of 0.90, an ROC-AUC of 0.97, and a PR-AUC of 0.97 on the test set. Feature importance analysis identifies the encounter-impulse parameter as the dominant predictor of the outcome. The classifier provides a computational speedup of approximately 74 compared with direct N-body simulations. We further evaluate the model using an independent dataset representing a globular-cluster environment, where approximately 20 per cent of planets retain HZ orbits. Despite the naturally imbalanced class distribution, the model achieves an accuracy of 0.87, an ROC-AUC of 0.92, and a PR-AUC of 0.77. These results demonstrate the potential of machine learning for efficiently predicting the dynamical retention of HZ planets following stellar encounters and for investigating their outcomes in dense stellar environments.
\end{abstract}




\section{Introduction}

Stars are frequently born in stellar groups and clusters, where the relatively
high stellar densities can expose planetary systems to gravitational
perturbations from nearby stars \citep{LadaLada2003,Li2019,Cai2018}. Although planetary systems are often studied as isolated systems \citep{BrownRein2022,Boley2012}, their dynamical evolution can be significantly influenced by their stellar birth environments \citep{Winter2020,vanElteren2019}. In particular, close stellar encounters, commonly referred to as stellar
flybys, can perturb planetary orbits, exciting eccentricities and
inclinations, changing orbital separations, triggering dynamical
instabilities, and, in sufficiently close interactions, leading to the
ejection of planets from their host systems \citep{Malmberg2011,FujiiHori2019}.

The outcome of a stellar flyby depends on several parameters,
including the masses of the host and perturbing stars, the encounter
distance, the relative velocity, the geometry of the encounter, and
the initial orbital properties of the planet
\citep{Malmberg2011,BreslauPfalzner2019}. Numerical investigations have shown that stellar encounters can produce a range of dynamical outcomes. A planet may remain bound to its original host but acquire a more eccentric or wider orbit \citep{Shara2016,vanElteren2019}; it may become unbound and escape the system; or, under certain conditions, it may be captured by the passing star \citep{Malmberg2011,Li2019,Raymond2024}. Furthermore, even encounters that do not immediately remove planets can perturb their orbits sufficiently to trigger subsequent dynamical evolution and, in multiplanetary systems, planet--planet scattering \citep{Malmberg2011,Li2019,BrownRein2022}. These results demonstrate that stellar flybys can influence not only the immediate configuration of planetary systems but also their subsequent dynamical evolution \citep{Malmberg2011,BrownRein2022}.

One important aspect of planetary-system evolution concerns planets orbiting within the habitable zone (HZ) of their host stars. The habitable zone is conventionally defined as the range of orbital distances within which a rocky planet with a suitable atmosphere could maintain liquid water on its surface \citep{Kasting1993,Kopparapu2013}. Its boundaries are determined primarily by the stellar luminosity and effective temperature, as well as by assumptions about the planetary atmosphere and climate \citep{Kasting1993,Kopparapu2013}. Since the location of the habitable zone depends on the properties of the host star \citep{RamirezKaltenegger2014}, its boundaries vary among stars of different masses and luminosities \citep{Kopparapu2013}. However, a planet's initial location within this region does not necessarily imply that it will remain there throughout the dynamical evolution of its system \citep{Winter2020,Raymond2024}.

A stellar flyby can perturb a planet's orbit and consequently change both its semimajor axis and orbital eccentricity \citep{Shara2016,SchoettlerOwen2024,JimenezTorres2021}. Therefore, evaluating the effect of an encounter solely through the final semimajor axis may not provide a complete description of the planet's orbital location relative to the habitable zone \citep{WilliamsPollard2002}. For an eccentric orbit, the planet moves between its periastron and apastron distances, which may extend beyond the habitable-zone boundaries even when its semimajor axis remains within that region \citep{Bolmont2016,WilliamsPollard2002,Eggl2013}. In this study, we therefore adopt a more restrictive orbital criterion: a planet is considered to retain a habitable-zone orbit only if it remains gravitationally bound to its original host star and its entire final orbital range, from periastron to apastron, lies within the original habitable-zone boundaries. This criterion allows us to distinguish between planets that remain bound to their host stars and those that also retain orbits entirely within the habitable zone.

In this work, we investigate the dynamical effects of stellar flybys on planets initially located within the habitable zone of their host stars. We perform direct N-body simulations of three-body systems consisting of a host star, an orbiting planet, and a passing star. We explore a broad range of host-star and perturber masses, planetary orbital properties, and encounter parameters to investigate how stellar flybys affect the final orbital configurations of habitable-zone planets. For each simulated encounter, we determine whether the planet
remains bound to its original host and whether its entire final
orbital range is contained within the habitable zone.

In addition to the numerical simulations, we develop a machine-learning model based on XGBoost \citep{ChenGuestrin2016} to predict whether a planet retains a habitable-zone orbit following a stellar encounter. The model uses the initial conditions and derived parameters
of the encounters to identify the combinations of physical properties
that are most informative for predicting the outcome. We evaluate its predictive performance on an independent test set
and further examine its applicability to simulated encounters under
globular-cluster conditions. This approach provides a means of predicting the retention of
habitable-zone orbits without requiring a new N-body integration
for every encounter.

The remainder of this paper is organized as follows.
Section~\ref{sec:method} describes the numerical simulations, including the adopted initial conditions, the determination of the maximum impact parameter, and the classification of the final orbital outcomes, followed by the machine-learning methodology and model training procedure.
Section~\ref{sec:results} presents the results of the machine-learning model, including its predictive performance, the relative importance of the input features, its application to simulated encounters in globular-cluster environments, and a comparison of its computational efficiency with direct N-body simulations.
Section~\ref{sec:discussion} discusses the implications of our findings, their limitations, and possible directions for future work.
Finally, Section~\ref{sec:conclusions} summarizes the main conclusions of this study.

\section{Method}
\label{sec:method}

\subsection{Numerical Simulations}
\label{subsec:numerical_simulations}

\subsubsection{Initial Conditions and Encounter Parameters}

We performed a large suite of three-body simulations to investigate
the dynamical effects of stellar flybys on planets initially residing
within the habitable zone (HZ) of their host stars. Each system
consists of a central star, a planet, and a passing star. All
simulations were carried out using the REBOUND N-body package
\citep{ReinLiu2012}, with calculations performed in units of solar
masses, astronomical units, and years.

The mass of the central star, $M_\star$, was sampled from a logarithmically uniform distribution between $0.1$ and $1.5\,M_\odot$ \citep{Kopparapu2013}. For each central star, its luminosity and radius were calculated using empirical mass-dependent relations \citep{Tout1996}. The effective temperature was then obtained from the Stefan--Boltzmann relation,

\begin{equation}
T_{\rm eff}=T_\odot
\left(\frac{L_\star}{R_\star^2}\right)^{1/4},
\end{equation}

where $T_\odot=5772\,{\rm K}$, and $L_\star$ and $R_\star$ are expressed in solar units.

The HZ boundaries were determined following the prescriptions
of \cite{Kopparapu2013,Kopparapu2014}, adopting the runaway greenhouse limit for the inner boundary and the maximum greenhouse limit for the outer boundary. The corresponding effective stellar fluxes were calculated as functions of the stellar effective temperature. For the inner boundary, the coefficients of the runaway greenhouse prescription were interpolated according to the planetary mass, while the maximum greenhouse prescription was adopted for the outer boundary \citep{Kopparapu2014}. The orbital distance of each boundary was calculated as

\begin{equation}
d_{\rm HZ}=\sqrt{\frac{L_\star}{S_{\rm eff}}},
\end{equation}

where $S_{\rm eff}$ is the corresponding effective stellar flux \citep{Kopparapu2014}.

Planetary masses, $M_{\rm p}$, were sampled logarithmically between $0.1$ and $5\,M_\oplus$ following the planetary-mass range
considered by \cite{Kopparapu2014}. The initial semimajor axis, $a_{\rm p}$, was drawn from a uniform distribution between the inner and outer HZ boundaries. The eccentricity was also sampled uniformly, with its maximum value chosen to ensure that the entire initial orbit remained within the HZ. Specifically, we required both the periastron and apastron distances to lie within the HZ boundaries, giving

\begin{equation}
e_{\max}=
\min\left[
1-\frac{d_{\rm HZ,in}}{a_{\rm p}},
\frac{d_{\rm HZ,out}}{a_{\rm p}}-1,
0.999
\right].
\end{equation}

The initial eccentricity was then drawn uniformly from $[0,e_{\max}]$. The initial true anomaly, $f_{\rm p}$, was randomly selected from $\{0,\pi\}$, corresponding to initial positions at periastron and apastron, respectively.

The passing-star mass, $M_3$, was sampled from a logarithmically uniform distribution over the range $0.1$--$150\,M_\odot$ \citep{Hopkins2018,WeidnerKroupa2004}. The initial separation between the passing star and the central star was set to

\begin{equation}
r_{\rm ini}=
a_{\rm p}\max\left[
10,\,
10\left(\frac{2M_3}{M_\star}\right)^{1/3}
\right].
\end{equation}

The initial relative speed of the passing star was required to exceed the two-body escape speed at this separation,

\begin{equation}
v_{\rm esc}=
\sqrt{\frac{2G(M_\star+M_3)}{r_{\rm ini}}},
\end{equation}

where $G$ is the gravitational constant. The incoming speed, $v$, was sampled logarithmically between $v_{\rm esc}$ and $10^3\,{\rm au\,yr^{-1}}$.

The impact parameter, $b$, was also sampled from a logarithmically uniform distribution. Its lower limit was set to $0.5a_{\rm p}$, while its upper limit was set to $1.5b_{\rm cutoff}$, where $b_{\rm cutoff}$ is given by the convergence-based fitting relation described in Section~\ref{subsec:impact_parameter_convergence}.

To account for different encounter geometries, the incoming direction of the passing star was sampled isotropically over the sphere. This was achieved by drawing $\cos\theta$ uniformly from $[-1,1]$ and the azimuthal angle $\phi$ uniformly from $[0,2\pi]$. The impact-parameter vector was constructed perpendicular to the incoming direction, with its orientation selected randomly in the perpendicular plane. The initial relative position and velocity vectors of the passing star were then assigned according to the sampled encounter parameters. Finally, the complete three-body system was transformed to its center-of-mass reference frame before integration.

The ranges and sampling distributions of the initial parameters are summarized in Table~\ref{tab:initial_conditions}.

\begin{table}
\caption{Initial parameter ranges and sampling distributions adopted in the numerical simulations.}
\label{tab:initial_conditions}
\centering
\begin{tabular}{lll}
\hline
Parameter & Range & Distribution \\
\hline
$M_\star$ & $[0.1,\,1.5]\,M_\odot$ & Log-uniform \\
$M_{\rm p}$ & $[0.1,\,5]\,M_\oplus$ & Log-uniform \\
$a_{\rm p}$ & $[d_{\rm HZ,in},\,d_{\rm HZ,out}]$ & Uniform \\
$e_{\rm p}$ & $[0,\,e_{\max}]$ & Uniform \\
$f_{\rm p}$ & $\{0,\,\pi\}$ & Discrete, equal probability \\
$M_3$ & $[0.1,\,150]\,M_\odot$ & Log-uniform \\
$r_{\rm ini}$ & $\displaystyle a_{\rm p}\max\left[10,\,10\left(\frac{2M_3}{M_\star}\right)^{1/3}\right]$ & Deterministic \\
$v$ & $[v_{\rm esc},\,10^3]$ au yr$^{-1}$ & Log-uniform \\
$b$ & $[0.5a_{\rm p},\,1.5b_{\rm cutoff}]$ & Log-uniform \\
$\cos\theta$ & $[-1,\,1]$ & Uniform \\
$\phi$ & $[0,\,2\pi]$ & Uniform \\
\hline
\end{tabular}
\end{table}

\subsubsection{Convergence Test for the Maximum Impact Parameter}
\label{subsec:impact_parameter_convergence}

We performed a series of convergence tests to determine an appropriate upper limit for the impact-parameter distribution. The tests covered three central-star masses, $M_\star=0.1,\ 0.8,$ and $1.5\,M_\odot$, and five passing-star masses, $M_3=0.1,\ 1,\ 10,\ 100,$ and $150\,M_\odot$. In all tests, the planetary mass was fixed at $2.55\,M_\oplus$, corresponding to the midpoint of the planetary-mass range adopted in the main simulations. The planet was initially placed on a circular orbit at the midpoint of the host star's HZ,

\begin{equation}
a_{\rm p}=\frac{d_{\rm HZ,in}+d_{\rm HZ,out}}{2}.
\end{equation}

For each mass combination, the search began at $b/a_{\rm p}=1.1$, and this ratio was increased by a factor of 1.2 at each subsequent step. At every tested impact parameter, 50 independent simulations were performed, with the encounter velocity and direction independently sampled. For these convergence tests, the initial separation was fixed at $r_{\rm ini}=100a_{\rm p}$, and the incoming speed was sampled logarithmically between $\max(v_{\rm esc},10^{-5}\,{\rm au\,yr^{-1}})$ and $10^3\,{\rm au\,yr^{-1}}$.

A simulation was considered to satisfy the HZ criterion if the planet remained within the HZ throughout the integration. This was assessed by recording the star--planet separation at $100{,}000$ equally spaced output times and checking whether

\begin{equation}
r_{\min}\geq d_{\rm HZ,in},
\qquad
r_{\max}\leq d_{\rm HZ,out},
\end{equation}

where $r_{\min}$ and $r_{\max}$ are the minimum and maximum recorded separations, respectively. A tested impact parameter was accepted if no more than five out of the 50 simulations failed to satisfy this condition.

To reduce the possibility of identifying a cutoff because of statistical fluctuations at a single impact parameter, we imposed an additional confirmation criterion. The next three consecutively tested, larger impact parameters were also required to meet the same acceptance condition. Thus, the cutoff was defined as the first value in a sequence of four consecutive accepted impact parameters. For a given central-star mass, the confirmed cutoff for one passing-star mass was used as the starting value for the search at the next passing-star mass. If no cutoff was confirmed within the tested range, the subsequent search was restarted from $b/a_{\rm p}=1.1$.

The confirmed cutoff values were used to obtain a fitting relation for the maximum impact parameter as a function of the central-star mass, passing-star mass, and planetary semimajor axis. The resulting relation is

\begin{equation}
b_{\rm cutoff}=
7.827\,M_\star^{-0.211}
M_3^{0.248}a_{\rm p},
\label{eq:impact_parameter_cutoff}
\end{equation}

where $M_\star$ and $M_3$ are expressed in solar masses and $a_{\rm p}$ is expressed in au. In the main simulations, the upper limit of the impact-parameter distribution was set to $1.5b_{\rm cutoff}$.

\subsubsection{Numerical Integration and Final Orbital Classification}

The dynamical evolution of each three-body system was followed by integrating the equations of motion using the IAS15 integrator implemented in REBOUND \citep{ReinLiu2012,ReinSpiegel2015}. The integration duration was set to three encounter timescales, where the encounter timescale was defined as

\begin{equation}
t_{\rm enc}=\frac{r_{\rm ini}}{v}.
\end{equation}

Accordingly, the total integration time was

\begin{equation}
t_{\rm sim}=3t_{\rm enc}.
\end{equation}

This integration interval allowed the passing star to travel through the system and move beyond the initial encounter region before the final orbital configuration was evaluated.

At the end of each integration, the planet's orbital elements were calculated relative to the central star. The final semimajor axis, $a_{\rm p,f}$, and eccentricity, $e_{\rm p,f}$, were used to determine whether the planet remained entirely within its original HZ. A planet was classified as remaining within the HZ if its final orbit satisfied

\begin{equation}
a_{\rm p,f}(1-e_{\rm p,f})>d_{\rm HZ,in}
\end{equation}

and

\begin{equation}
a_{\rm p,f}(1+e_{\rm p,f})<d_{\rm HZ,out}.
\end{equation}

\begin{figure*}
    \centering

    \begin{subfigure}[b]{0.48\textwidth}
        \centering
        \includegraphics[width=\textwidth]{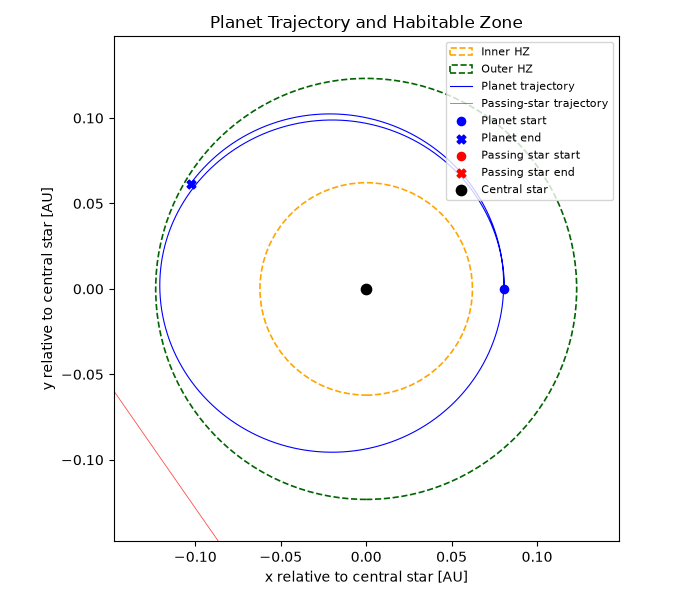}
        \caption{The planet remains within the habitable zone.}
        \label{fig:HZ_retained}
    \end{subfigure}
    \hfill
    \begin{subfigure}[b]{0.48\textwidth}
        \centering
        \includegraphics[width=\textwidth]{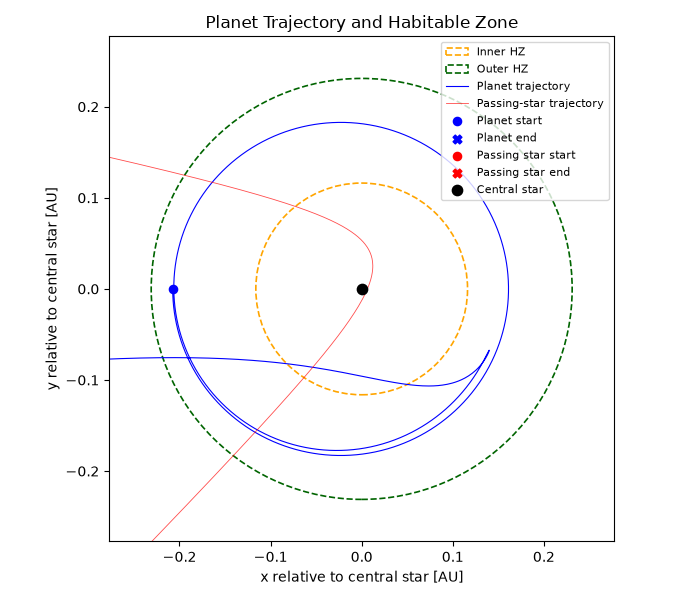}
        \caption{The planet moves beyond the habitable zone.}
        \label{fig:HZ_outside}
    \end{subfigure}

    \caption{Examples of the dynamical evolution of planetary orbits following stellar flyby encounters. The two panels show representative simulations in which the planet (a) remains entirely within the habitable zone (HZ) throughout the encounter and (b) moves beyond the HZ boundaries. The central star is shown in yellow, the inner and outer HZ boundaries in orange and green, respectively, the planetary trajectory in blue, and the passing star's trajectory in red. The shaded region indicates the habitable zone.}
    \label{fig:HZ_examples}

\end{figure*}

In addition, the final semimajor axis and eccentricity were required to be finite, with $0\leq e_{\rm p,f}<1$. Systems that did not satisfy all these conditions were classified as HZ-exiting systems. This criterion ensures that the entire final orbit, rather than only the planet's instantaneous position, lies within the original HZ. Figure~\ref{fig:HZ_examples} illustrates two representative outcomes of
stellar flyby encounters. In both panels, the passing star follows the
trajectory shown in red, gravitationally perturbing the planet's orbit.
While the planet in panel (a) remains entirely within the HZ by the end
of the simulation, the planet in panel (b) is displaced beyond the HZ
boundaries as a result of the stellar encounter.

As a numerical accuracy check, the total energy of the three-body system was calculated before and after each integration. The relative energy error was recorded as

\begin{equation}
\epsilon_E=
\left|
\frac{E_{\rm final}-E_{\rm initial}}
{E_{\rm initial}}
\right|.
\end{equation}

\subsubsection{Parallel Simulations and Data Collection}

To construct a large dataset of independent stellar encounters, the simulations were executed in parallel using Python's process-based parallel execution framework. A total of $200{,}000$ independent simulations were scheduled, with the number of parallel workers set to one fewer than the number of available CPU cores.

For each simulation, the output included the initial stellar and planetary parameters, the properties of the stellar encounter, and the final orbital elements of the planet. The initial and final positions and velocities of all three bodies were also recorded. Each simulation was assigned a unique identifier to facilitate the tracking and handling of individual runs.

The simulation results were written to CSV files, with each file
containing up to 50,000 successfully completed simulation records. Failed runs were excluded from the dataset, while their number and corresponding error messages were recorded separately during execution.

\subsection{Machine Learning}

We employed the XGBoost (Extreme Gradient Boosting) \citep{ChenGuestrin2016} algorithm
to develop a binary classifier that predicts whether a planet
remains within the habitable zone (HZ) following a stellar
flyby. The classifier was trained on the results of the
three-body simulations described in the previous section.
The target variable, \texttt{is\_HZ}, indicates whether the
planet's final orbit satisfies the adopted HZ-retention
criterion. Accordingly, the problem was formulated as a
binary classification task, with the two classes representing
planets that either remain within or leave the habitable zone.

\subsubsection{Feature Construction}

The initial input dataset consisted of 14 parameters describing
the planetary system, the passing star, and the encounter
geometry. These included the masses of the central star,
planet, and passing star; the planet's initial semimajor axis,
eccentricity, and true anomaly; the inner and outer boundaries
of the habitable zone; the relative velocity and impact
parameter of the encounter; the initial separation between
the stars; the encounter orientation angles; and the encounter
timescale.

In addition to these original simulation parameters, we
constructed a set of derived physical features to provide the
classifier with information about the relative scales and
dynamical characteristics of the encounter. These features
were designed to represent relevant combinations of the
system parameters rather than relying exclusively on their
individual values. The derived quantities are summarized in
Table~\ref{tab:physical_features}.

The stellar mass ratio, \(M_\star/M_3\), characterizes the
relative masses of the central and passing stars, while the
impact parameter ratio, \(b/a_{\mathrm{planet}}\), describes
the encounter distance relative to the planet's orbital scale.
The gravitational focusing parameter,
\(M_3/(bv^2)\), combines the perturber's mass, encounter
distance, and relative velocity, providing a measure of the
importance of gravitational deflection during the encounter.

The planet’s initial position within the habitable zone was
represented by the normalized HZ position,
\begin{equation}
\frac{a_{\rm planet}-d_{\rm HZ,in}}
{d_{\rm HZ,out}-d_{\rm HZ,in}},
\end{equation}
which describes its location relative to the inner and outer
HZ boundaries. The HZ width was calculated as an intermediate
quantity for this normalization but was not retained as an
independent input feature.

We also included an impulse-related parameter,
\(t_{\mathrm{enc}}M_3/b^2\), to combine the perturber's mass
and encounter timescale with the encounter distance. The
ratio between the encounter timescale and the characteristic
orbital timescale,
\begin{equation}
t_{\rm enc}\sqrt{\frac{a_{\rm planet}^3}{M_\star}}
\end{equation}
provides information about the duration of the encounter
relative to the planet's orbital timescale. Finally, the
quantity \(ba_{\mathrm{planet}}/r_{\mathrm{ini}}\) relates
the impact parameter and planetary orbital scale to the
initial separation of the two stars.

Following feature construction, the original parameters
\(M_\star\), \(M_3\), \(b\), \(a_{\mathrm{planet}}\),
\(d_{\mathrm{HZ,in}}\), \(d_{\mathrm{HZ,out}}\), and
\(f_{\mathrm{planet}}\), together with the intermediate HZ
width, were removed from the feature matrix. The remaining
original parameters were retained alongside the seven
derived features. Thus, the final input matrix consisted of
14 features per simulation. The resulting feature values
were converted to 32-bit floating-point format before being
passed to the classifier.

\begin{table}
    \centering
    \caption{Derived physical features constructed for the
    XGBoost classifier. The HZ width is used as an intermediate
    quantity in calculating the normalized HZ position and is
    not included independently in the final feature matrix.}
    \label{tab:physical_features}
    \renewcommand{\arraystretch}{1.2}
    \begin{tabular}{lc}
        \hline
        Feature & Definition \\
        \hline
        Stellar mass ratio
        & $M_\star/M_3$ \\

        Impact parameter ratio
        & $b/a_{\mathrm{planet}}$ \\

        Gravitational focusing
        & $M_3/(bv^2)$ \\

        HZ position
        & $\displaystyle
        \frac{a_{\mathrm{planet}}-d_{\mathrm{HZ,in}}}
        {d_{\mathrm{HZ,out}}-d_{\mathrm{HZ,in}}}$ \\

        Impulse-related parameter
        & $t_{\mathrm{enc}}M_3/b^2$ \\

        Encounter-to-orbital-timescale ratio
        & $\displaystyle
        \frac{t_{\mathrm{enc}}}
        {\sqrt{a_{\mathrm{planet}}^3/M_\star}}$ \\

        Impact-to-initial-distance ratio
        & $ba_{\mathrm{planet}}/r_{\mathrm{ini}}$ \\
        \hline
    \end{tabular}
\end{table}

\subsubsection{Hyperparameter Optimization and Model Training}

The XGBoost classifier was optimized using Bayesian
hyperparameter optimization \citep{Kenny2024,Brochu2010} with the Optuna framework and
its Tree-structured Parzen Estimator (TPE) sampler \citep{Akiba2019}. A total
of 100 optimization trials were performed to identify a
combination of hyperparameters that maximized the average
precision, or area under the precision--recall curve
(PR-AUC), on a validation subset. For this optimization
stage, the dataset was divided into training and validation
subsets using an 80:20 stratified split. Stratification
preserved the relative proportions of the two target classes
in both subsets. A fixed random state was used to ensure
reproducibility, and cross-validation was not employed during
hyperparameter optimization.

After optimization, the selected hyperparameters were used
to train the final classifier. The dataset was independently
divided into training and test subsets, again using an 80:20
stratified split. The model was fitted to the training subset,
while the test subset was reserved for the final performance
evaluation. The optimized hyperparameter values are listed
in Table~\ref{tab:optimized_hyperparameters}.

The final model consisted of 1600 boosting iterations, with
a maximum tree depth of 7 and a learning rate of 0.0190.
The subsampling and column-sampling fractions were set to
0.6444 and 0.9736, respectively, controlling the fractions
of training samples and input features considered during
tree construction. The minimum child weight and split
regularization parameter (\(\gamma\)) were set to 7.0024
and 3.2815, respectively. The L1 and L2 regularization
coefficients were 0.0008 and 0.0205, while the maximum
number of bins used for continuous feature discretization
was set to 512. The classifier used the binary logistic
objective function, with log-loss as its training evaluation
metric.

\begin{table}
    \centering
    \caption{Optimized hyperparameters adopted for the final
    XGBoost classifier.}
    \label{tab:optimized_hyperparameters}
    \renewcommand{\arraystretch}{1.15}
    \begin{tabular}{lc}
        \hline
        Hyperparameter & Optimized Value \\
        \hline
        n\_estimators       & 1600 \\
        max\_depth          & 7 \\
        learning\_rate      & 0.0190 \\
        subsample           & 0.6444 \\
        colsample\_bytree   & 0.9736 \\
        min\_child\_weight  & 7.0024 \\
        gamma               & 3.2815 \\
        reg\_alpha          & 0.0008 \\
        reg\_lambda         & 0.0205 \\
        max\_bin            & 512 \\
        \hline
    \end{tabular}
\end{table}

To further assess the robustness of the trained classifier,
we performed a 10-fold stratified cross-validation on the
training subset. The average precision (PR-AUC) was used
as the scoring metric for each fold. This cross-validation
was conducted after hyperparameter optimization and was
separate from the optimization procedure itself.

\subsubsection{Model Evaluation}

The trained classifier was evaluated on the held-out test
subset. Class predictions were obtained using the default
classification threshold of 0.5, while the predicted
probabilities were used for probability-based evaluation
metrics. We quantified classification performance using
accuracy, precision, recall, specificity, and the F1-score.
In addition, the area under the receiver operating
characteristic curve (ROC-AUC) and the area under the
precision--recall curve (PR-AUC) were calculated to assess
the model's ability to distinguish between the two classes.
The quality of the predicted probabilities was further
examined using the Brier score and expected calibration
error (ECE).

In addition to the held-out test evaluation, an independent
dataset of REBOUND simulations, excluded from the training
dataset, was used to assess the classifier on a separate
set of simulations. The same feature-construction procedure
was applied to this dataset, and the trained model was used
to predict the HZ-retention outcome. Its predictions were
then compared with the corresponding simulation labels
using the same evaluation metrics.

\section{Results}
\label{sec:results}

\subsection{Classification Performance}

\begin{figure}
    \centering
    \includegraphics[width=0.9\columnwidth]{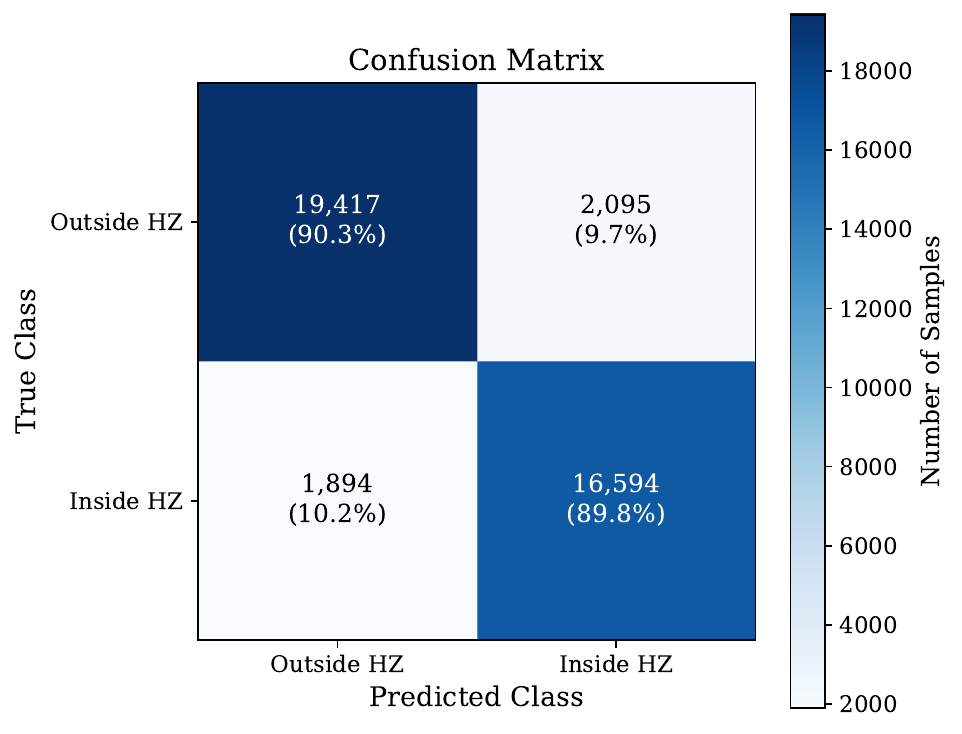}
    \caption{Confusion matrix of the XGBoost classifier evaluated
    on the held-out test set of 40,000 samples. The rows represent
    the actual classes, while the columns indicate the predicted
    classes. The diagonal entries correspond to correctly
    classified samples: 19,417 true negatives (Not\_HZ) and
    16,594 true positives (HZ). The off-diagonal entries show
    the misclassified samples, comprising 2,095 false positives
    and 1,894 false negatives.}
    \label{fig:confusion_matrix}
\end{figure}

The performance of the XGBoost classifier was evaluated on a
held-out test set comprising 40,000 samples. The confusion
matrix in Figure~\ref{fig:confusion_matrix} shows that the
classifier correctly identified 19,417 samples belonging to
the Not\_HZ class (true negatives) and 16,594 samples belonging
to the HZ class (true positives). It incorrectly classified
2,095 Not\_HZ samples as HZ (false positives) and 1,894 HZ
samples as Not\_HZ (false negatives). These results correspond
to an overall accuracy of 90.03\%, indicating that the model
correctly classified the large majority of the test samples.

The model achieved a precision of 0.8879, a recall of 0.8976,
and an F1-score of 0.8927. Its specificity was 0.9026,
indicating a similar ability to identify planets that leave
the habitable zone and those that remain within it. The
difference between recall and specificity is small, suggesting
that the classifier does not exhibit a strong preference
towards either class at the adopted classification threshold.
The ROC-AUC and PR-AUC values were 0.9708 and 0.9681,
respectively, indicating strong discriminative performance.
In addition to classification performance, the probabilistic
predictions were evaluated using the Brier score and expected
calibration error (ECE), which yielded values of 0.0689 and
0.0341, respectively.

To examine the consistency of the model across different
subsets of the training data, we also performed a 10-fold
stratified cross-validation \citep{Hastie2009,LiuZhou2024}. The mean PR-AUC across the
10 folds was \(0.9687 \pm 0.0010\), indicating that the
model achieved consistent performance across the different
folds. The test-set metrics and cross-validation result are
summarized in Table~\ref{tab:xgboost_performance}.

\begin{table}
\centering
\small
\caption{Performance metrics of the final XGBoost classifier.
The metrics were calculated on the held-out test set of
40,000 samples. The final row reports the mean and standard
deviation of the average precision (PR-AUC) obtained from
10-fold stratified cross-validation on the training subset.}
\label{tab:xgboost_performance}
\begin{tabular}{lc}
\hline
Metric & Value \\
\hline
Accuracy              & 0.9003 \\
Precision             & 0.8879 \\
Recall                & 0.8976 \\
Specificity           & 0.9026 \\
F1-score              & 0.8927 \\
ROC-AUC               & 0.9708 \\
PR-AUC                & 0.9681 \\
Brier score           & 0.0689 \\
Expected Calibration Error (ECE) & 0.0341 \\
10-fold CV PR-AUC     & \(0.9687 \pm 0.0010\) \\
\hline
\end{tabular}
\end{table}

\subsection{Feature Importance}

\begin{figure}
    \centering
    \includegraphics[width=0.9\columnwidth]{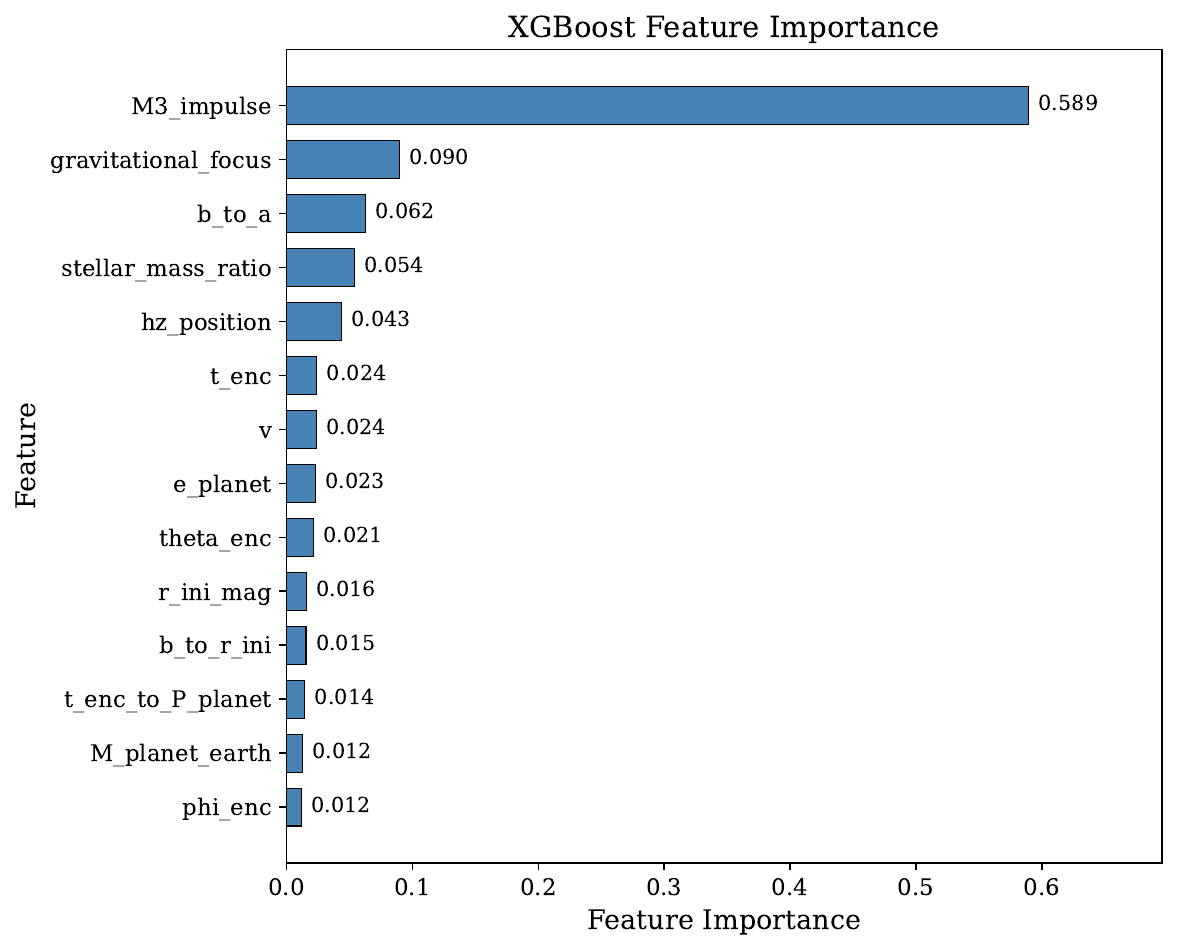}
    \caption{XGBoost feature importance scores for the 14 input
    features used by the classifier. The scores indicate the
    relative contributions of the features to the model's
    predictions, with larger values corresponding to greater
    importance within the trained model. The \(M3\_impulse\)
    feature has the largest importance score, followed by
    \(gravitational\_focus\), \(b\_to\_a\), and
    \(stellar\_mass\_ratio\). The scores represent model-based
    feature importance and do not directly measure causal
    physical effects.}
    \label{fig:feature_importance}
\end{figure}

The relative contributions of the 14 input features to the
model predictions were examined using the feature importance
scores provided by XGBoost. The resulting importance
distribution is shown in Figure~\ref{fig:feature_importance}.
Among all input features, \(M3\_impulse\) was by far the most
important, with an importance score of 0.5894. Its contribution
was substantially higher than that of the second-ranked
feature, \(gravitational\_focus\), which had a score of 0.0897.

The next most important features were \(b\_to\_a\) (0.0624),
\(stellar\_mass\_ratio\) (0.0539), and \(hz\_position\)
(0.0433). Together, these five features accounted for
approximately 83.9\% of the total feature importance. The
remaining nine features had considerably smaller importance
scores, each contributing less than 0.025. In particular,
\(M\_planet\_earth\) and \(\phi_{\mathrm{enc}}\) had the
lowest scores, at 0.0124 and 0.0118, respectively.

The dominance of \(M3\_impulse\) indicates that this derived
parameter provided the largest contribution to the
classifier's predictions among the features considered.
The relatively high importance of \(gravitational\_focus\),
\(b\_to\_a\), and \(stellar\_mass\_ratio\) further highlights
the predictive relevance of the encounter geometry and the
relative dynamical properties of the interacting stars.
In contrast, the lower importance assigned to the planetary
mass suggests that, within the parameter ranges explored in
this study, it provided less predictive information than
the features describing the stellar encounter. These
importance scores, however, quantify the contribution of
features to the trained model and should not be interpreted
as direct measures of causal physical influence.

\subsection{Model Evaluation in Globular-Cluster Environments}

We further evaluated the trained XGBoost model in a
globular-cluster environment using a new dataset generated
from simulated stellar encounters. The full cluster dataset
reflects the natural class distribution in this environment,
with approximately 20\% of planets remaining within the
habitable zone. To examine the effect of class imbalance on
the evaluation metrics, we considered both the full dataset
and a balanced subset obtained by randomly undersampling
the majority class.

Table~\ref{tab:cluster_evaluation} summarizes the model's
performance on these two datasets. On the full dataset, the
model achieved an accuracy of 0.8738, a precision of 0.6584,
a recall of 0.7221, and an F1-score of 0.6888. The ROC-AUC
and PR-AUC were 0.9158 and 0.7702, respectively. Following
random undersampling of the majority class, the precision
increased to 0.8901 and the F1-score to 0.7974, while the
PR-AUC increased to 0.9174. The recall remained unchanged
at 0.7221, and the ROC-AUC was also nearly identical for
the two datasets.

The differences in precision and PR-AUC between the full
and balanced datasets should be interpreted in the context
of their different class distributions. Both metrics are
sensitive to the prevalence of the positive class, and
therefore their values are not directly comparable across
datasets with different class proportions \citep{Saito2015}. In contrast,
the similar ROC-AUC values indicate that the model retained
comparable ranking performance under the two class
distributions. The results therefore provide an assessment
of the classifier under the naturally imbalanced cluster
distribution as well as under a balanced evaluation sample.

\begin{table}
\centering
\small
\caption{Performance of the trained XGBoost classifier on
the globular-cluster dataset. Results are reported for the
full dataset, which retains the naturally imbalanced class
distribution, and for a balanced subset obtained by randomly
undersampling the majority class. The metrics quantify the
classifier's performance under the two different class
distributions.}
\label{tab:cluster_evaluation}
\begin{tabular}{lcc}
\hline
Metric & Full dataset & Balanced dataset \\
\hline
Accuracy      & 0.873760 & 0.816481 \\
Precision     & 0.658356 & 0.890085 \\
Recall        & 0.722136 & 0.722136 \\
Specificity   & 0.910124 & 0.910825 \\
F1-score      & 0.688773 & 0.797363 \\
ROC-AUC       & 0.915815 & 0.916318 \\
PR-AUC        & 0.770173 & 0.917402 \\
Brier score   & 0.090217 & 0.132354 \\
ECE           & 0.044899 & 0.094956 \\
\hline
\end{tabular}
\end{table}

\subsection{Computational Speed Benchmark}

The computational throughput of the trained XGBoost model
was compared with that of direct N-body simulations performed
using REBOUND. To make the comparison consistent, the
computational throughput of both approaches was measured
using a single CPU core. Under these conditions, REBOUND
completed approximately 657 simulations per second, whereas
the trained machine-learning model performed approximately
48,820 predictions per second. Thus, the machine-learning
approach achieved a speedup of approximately 74 times
relative to direct N-body simulations.

For example, processing 50,000 stellar encounters would
require approximately 76 seconds using direct N-body
simulations, compared with just over 1 second using the
trained classifier. This difference illustrates the
computational advantage of using the trained model to
estimate the outcomes of stellar encounters once the
initial training simulations have been completed. In
particular, the classifier can facilitate the exploration
of large parameter spaces and extensive statistical
investigations for which running a large number of direct
N-body simulations would be computationally more demanding.

\section{Discussion}
\label{sec:discussion}

The results of this study demonstrate that stellar flybys
can alter the orbital configurations of planets initially
located within the habitable zones of their host stars.
The extent of this effect depends on the physical and
orbital properties of the interacting system, including
the masses of the host and passing stars, the encounter
distance, and the initial planetary orbit \citep{Malmberg2011,FlamminiDotti2019}. In particular,
the adopted HZ-retention criterion provides a more
restrictive assessment of the dynamical outcome than
simply determining whether a planet remains gravitationally
bound to its host. A planet may survive an encounter without
being ejected, yet its orbital eccentricity or semimajor
axis may change sufficiently for part or all of its orbit
to extend beyond the habitable-zone boundaries \citep{Malmberg2011,LiMustillDavies2020,JimenezTorres2013}. Therefore,
distinguishing between gravitational retention and
retention within the habitable zone is important when
assessing the potential dynamical consequences of stellar
encounters for planetary systems \citep{Malmberg2011,BojnordiArbab2021}.

The feature importance analysis provides further insight
into the parameters that contribute most strongly to
predicting these outcomes. Among the 14 input features,
\(M3\_impulse\) dominates the model predictions, accounting
for approximately 58.9\% of the total feature importance.
This is followed by gravitational focusing, the impact
parameter relative to the planetary semimajor axis, and
the stellar mass ratio. The prominence of these features
highlights the predictive importance of the dynamical
characteristics of the encounter, particularly those
associated with the strength of the gravitational
perturbation and the relative properties of the interacting
stars \citep{Malmberg2011}. In contrast, the relatively low importance assigned
to the planetary mass suggests that, within the parameter
ranges explored in this study, the properties of the
stellar encounter provide more predictive information
than the planet's mass \citep{Li2020,Maraboli2023}. However, feature importance scores
quantify the relative contribution of individual features
to the model predictions and should not be interpreted
as direct measures of their causal physical effects \citep{Molnar2022}.
Moreover, because some of the derived features encode
related aspects of the encounter, their individual
importance scores should be interpreted in the context
of the complete feature set \citep{Molnar2022}.

Beyond its performance on the original simulation dataset,
the trained model was also evaluated in a globular-cluster
environment. This provides an additional assessment of
its predictive capability under a different set of
encounter conditions and a naturally imbalanced class
distribution. Approximately 20\% of the planets in the
full cluster dataset remain within the habitable zone.
Despite this imbalance, the model achieves an accuracy
of 0.87, an ROC-AUC of 0.92, and a recall of 0.72. Its
precision of 0.66 and PR-AUC of 0.77 are lower than those
obtained for the balanced subset. After random
undersampling of the majority class, the precision and
PR-AUC increase to 0.89 and 0.92, respectively, while
the recall remains unchanged. The similar ROC-AUC values
obtained for the two datasets indicate that the model
retains comparable ranking performance under both class
distributions \citep{Saito2015}. The increase in precision and PR-AUC after
balancing should not, however, be interpreted as an
improvement in the classifier itself, since these metrics
are affected by the prevalence of the positive class \citep{Saito2015}.
Furthermore, the Brier score and ECE are higher for the
balanced subset, indicating that the probabilistic
predictions are less well calibrated under this
evaluation distribution. These differences emphasize
the importance of considering class prevalence \citep{SaitoRehmsmeier2015} and
probability calibration \citep{NiculescuMizilCaruana2005} when applying the classifier
to stellar environments with different population
distributions. Overall, the results suggest that the
trained classifier can provide useful predictions when
applied to stellar encounters in a globular-cluster
environment beyond the original simulation dataset.

Several limitations of the present study should also
be considered. Our simulations focus on isolated
three-body encounters involving a single planet, its
host star, and one passing star. Consequently, the
results do not account for the combined effects of
multiple planets, repeated stellar encounters, or the
long-term dynamical evolution of planetary systems
within their host clusters. Furthermore, our
classification criterion is based on whether the final
planetary orbit lies entirely within the adopted
habitable-zone boundaries. It does not incorporate
atmospheric evolution, planetary climate, or other
conditions required to establish actual surface
habitability \citep{Kasting1993}. The present framework therefore evaluates
the dynamical retention of planets within the habitable
zone rather than their long-term habitability. Future
work could extend this approach to multiplanetary
systems and repeated stellar encounters, as well as
explore a wider range of stellar environments and
encounter histories. Such extensions would help
establish a more comprehensive picture of the role
of stellar dynamics in shaping the orbital evolution
of potentially habitable planets.

\section{Summary and Conclusions}
\label{sec:conclusions}

The main steps and findings of this study can be summarized as follows:

\begin{itemize}

    \item We performed a large suite of 140,000 three-body N-body simulations using REBOUND to investigate the dynamical effects of stellar flybys on planets initially located within the habitable zones of their host stars. The simulations covered a broad range of stellar masses, planetary orbital properties, and encounter parameters.

    \item We adopted an HZ-retention criterion based on the final periastron and apastron distances of the planet. A planet was classified as retaining an HZ orbit only if it remained gravitationally bound to its original host star and its entire final orbit was contained within the HZ boundaries. This criterion allowed us to distinguish between gravitational survival and the retention of an orbit within the habitable zone.

    \item Using the simulation dataset, we constructed 14 input features describing the physical properties and dynamical characteristics of the planetary systems and stellar encounters. We then developed an XGBoost classifier, with its hyperparameters optimized through Bayesian optimization using the Optuna framework.

    \item On the held-out test set of 40,000 samples, the classifier achieved an accuracy of 0.9003, an ROC-AUC of 0.9708, and a PR-AUC of 0.9681. A 10-fold stratified cross-validation further yielded a mean PR-AUC of \(0.9687 \pm 0.0010\), indicating consistent predictive performance.

    \item The feature importance analysis identified \(M3\_impulse\) as the dominant predictor, accounting for approximately 58.9\% of the total feature importance. Gravitational focusing, the impact parameter relative to the planetary semimajor axis, and the stellar mass ratio were the next most important features. Together, the five most important features accounted for approximately 83.9\% of the total importance.

    \item We further evaluated the classifier using a dataset representing stellar encounters in a globular-cluster environment, where approximately 20\% of planets retained HZ orbits. On the full, naturally imbalanced dataset, the model achieved an accuracy of 0.8738 and an ROC-AUC of 0.9158, with a PR-AUC of 0.7702. On a balanced subset, the F1-score and PR-AUC increased to 0.7974 and 0.9174, respectively, while the ROC-AUC remained nearly unchanged.

    \item Finally, we compared the computational throughput of the trained classifier with that of direct N-body simulations using a single CPU core. The classifier performed approximately 48,820 predictions per second, compared with 657 simulations per second using REBOUND, corresponding to a speedup of approximately 74 times.

\end{itemize}

Overall, our results demonstrate that stellar flybys can substantially modify the orbital configurations of planets initially located within the habitable zones of their host stars. Importantly, a planet may remain gravitationally bound after an encounter while no longer retaining an orbit entirely within the habitable zone. The developed XGBoost classifier provides an efficient means of predicting this dynamical outcome, with the properties of the stellar encounter playing a more prominent role in the model predictions than the planetary mass within the parameter ranges explored. Its evaluation in a globular-cluster environment further demonstrates its potential applicability to stellar populations with different encounter conditions, although the effects of class imbalance and probability calibration must be taken into account. By providing predictions considerably faster than direct N-body integrations, the model can facilitate the exploration of large encounter parameter spaces and the statistical investigation of planetary systems in dense stellar environments. Nevertheless, the present study is limited to isolated three-body encounters involving a single planet and does not account for multiple planets, repeated stellar encounters, or the long-term dynamical evolution of planetary systems within clusters. Moreover, the adopted criterion describes the dynamical retention of planetary orbits within the habitable zone rather than actual surface habitability, which also depends on atmospheric and climatic conditions. Future studies incorporating multiplanetary systems, repeated encounters, and a broader range of stellar environments will help provide a more comprehensive understanding of the role of stellar dynamics in shaping the orbital evolution of potentially habitable planets.

\section*{Acknowledgements}

We acknowledge the use of ChatGPT (OpenAI) for assistance with code development and debugging, as well as language refinement and editing during the preparation of this manuscript. All AI-assisted outputs were reviewed and verified by the authors, who take full responsibility for the content of this work.




\bibliographystyle{mnras}
\bibliography{example} 





\bsp	
\label{lastpage}
\end{document}